# Generating Cycloidal Gears for 3D Printing

Sunny Daniels
e-mail: [sdaniels@lycos.com](mailto:sdaniels@lycos.com)

## Abstract

This article describes an algorithm for producing, for any desired resolution and any desired numbers of wheel and pinion teeth, polygonal approximations to the shapes of a pair of cycloidal gears that mesh correctly.  The larger (with the larger number of teeth) of the two gears is called the *wheel*; the smaller of the two gears is called the *pinion*.  An Octave implementation of the algorithm, mostly written in 2014, is included.  The Octave implementation contains a (crude, but evidently adequate, at least for reasonable numbers of wheel and pinion teeth) solution of the problem of finding (iteratively, since I am not aware of any analytical solution) the generating wheel angle corresponding to the tips of the tooth addenda.  (We'll call this the "tooth tip problem").  The only previous solution to this problem of which I am aware is somewhat more complicated.

However, this Octave implementation does not contain a good solution to the problem (not discussed in the literature at all, as far as I am aware) of automatically determining the generating wheel angles required to produce a polygon which approximates the curved parts of the teeth (the addenda) to a resolution specified by the user (in order to match the resolution of the 3D printer that will be used to physically produce the gears).  I was unsure as to how to solve this latter problem (we'll call this the "generating wheel step problem") until about six months ago, when I realised that there is (if I am not mistaken) a simple solution involving a priority queue.  A sketch of this proposed solution is included here.  However, since implementing a priority queue in Octave would be a bit cumbersome, my intention is to port the implementation to Java (which seems to have support for priority queues in its standard class libraries), and *then* attempt to implement the algorithm described here for automatically determining the generating wheel angles.

## Introduction

I first became interested in designing my own solar-thermal power station in November 2008, when, in the process of trying to help the landlady troubleshoot the solar (photovoltaic) power system in the house in which I was staying (in Pamapuria, near Kaitaia, here in New Zealand), I looked on the Internet for information on solar cells.  When, if I remember rightly, I saw the Wikipedia article on solar power, I was amazed to learn of the enormous amount of optical power (>1000 watts per square metre peak at midday on the equator on the equinox, about 200 watts per square metre global average, averaged over 24 hours per day and 365 days per year) in direct sunlight!

As far as I am aware, the amount of electricity that could be generated, using all of the World's land, with solar-thermal power (using concentrated sunlight to power a heat engine, as with the recently-completed Ivanpah station in California (reference [1]): *not* photovoltaic power) exceeds the amount of electricity that is currently being generated around the World *from all sources* by a factor of at least a hundred.

However, I believe that, historically, the problem with solar-thermal power has been the cost of producing the optics for concentrating the sunlight (heliostats in the case of Ivanpah), and the cost of producing the mechanism for moving these optical components to allow for the motion of the sun.

I wrote up drafts of a few ideas in this field between 2008 and 2013 but I was unaware of the existence of cheap 3D printers until late 2013. In late 2013 I was introduced to the world of 3D printing by a chance encounter with a programmer, Andrew Dixon, here in Auckland. I immediately thought of the possibility of designing my own mechanism for moving the heliostats from plastic, with the idea that it could be ultimately mass-produced very cheaply with plastic moulding equipment.

My idea was to gear down the output of a cheap, readily-available electric motor (for example, an old microwave oven fan motor) with several stages of plastic gearing, to the point where the output has enough torque to turn a heliostat mirror of reasonable size. (A mirror 50cm by 50cm can reflect enough sunlight onto a target to generate about NZ$100 worth of electricity per year, if I remember my calculations correctly: sorry, it is now a while since I did the calculations).

I started trying to design the gearing mechanism with OpenSCAD (reference [2]), but very quickly realised that:

1) Designing gears is *not* simple!

2) I knew nothing about designing gears at the time: I am a Computer Science and Mathematics graduate, not a mechanical engineer.

Hence, I set about the task of learning how to design my own gears for the mechanism to drive the heliostats. In the process, I believe I devised a complete algorithm for production of cycloidal gears, and made some progress toward devising robust solutions to two important problems that arise in the production of cycloidal gears: I am not aware of solutions to these two problems having been previously clearly described in the literature. (These are the "tooth tip problem" and "generating wheel step problem" that we discussed in the abstract).

## **Professor Faydor Litvin and his Textbook:**

I discovered farily quickly, by Google searching, that Professor Faydor Litvin, apparently originally from the Soviet Union, but now at the University of Illinois at Chicago, appears to be generally regarded as the world expert in gears. His 100th birthday was in 2014 (reference [3]).

His textbook "gear geometry and applied theory" (reference [4]) seems to be highly regarded and widely read. This textbook appears to have evolved from a NASA technical report published by Professor Litvin in 1989 (reference [5]).

Unfortunately I do not have easy access to Litvin's book here in the Auckland City Centre: the University of Auckland Library Catalogue says that there is only one copy of it in the Auckland library system (in the Engineering library), and says that it has been lost. There does not appear to be any copy of Litvin's book in the AUT library system.

## **Involute Gearing:**

It appears that the most common (by far) type of gear used in machinery is the involute gear. I believe that most car gearboxes use involute gearing, although I think a few use Wildhaber-Novikov gearing: It would appear from the contents of Litvin's book that he describes Wildhaber-Novikov gearing in chapter 17 (reference [4]). I believe that the advantage of Wildhaber-Novikov gearing over involute gearing is higher power density.

It appears from his table of contents that Litvin first describes involute gearing in chapter 10 of his book (reference [4]). Apparently involute gears were invented by Leonhard Euler (I don't rememeber exactly where I read this). As far as I know, both sides of a tooth of an involute gear are involutes of a circle.

There did seem to be freely available software for generating involute gears (in 2014), but evidently no such software that outputs OpenSCAD source files. It quickly became apparent, however, that the number of parameters that the user must specify to define a pair of meshing involute gears (inputs to the gear generator) is generally greater than that for a pair of meshing cycloidal gears (although this depends a bit on how many values are hard-coded into the gear generator): there is always at least the number of pinion teeth, number of wheel teeth and pressure angle for a pair of meshing involute gears, together with a few clearance values.

On the other hand, for a pair of meshing cycloidal gears, there is basically just the number of wheel teeth and number of pinion teeth.  The apparent greater complexity of involute gears (for the *user* of a gear-generating program) gave me a strong incentive to prefer cycloidal gearing for my application.

## **Cycloidal Gearing:**

I clearly remember first seeing Hugo Sparks's web page on cycloidal gearing (reference [6]) in late 2013 or early 2014.  I immediately liked the apparent simplicity of cycloidal gearing (similicity of the *inputs* to the gear generating software, that is; a major historical disadvantage of cycloidal gearing would appear to be the fact that some of the numerical quantities required to draw the gear teeth cannot be computed analytically; however, in today's world, where even US$1 microcontrollers have enough computing power to iteratively calculate these quantities in a fraction of a second, this disadvantage seems to be almost irrelevant).  **I first learned the meanings of the terms *hypocycloid*, *epicycloid, pitch circle, module, addendum* and *dedendum* from Hugo Sparks's web page.  Please read through his web page carefully if you are not familiar with all of these terms.**

I initially implemented Hugo Sparks' gear constuction in C, including his iterative method of calculating the quantities $\theta$ and $\beta$ that satisfy his equations (9) and (11).  (The reason that I used C rather than Octave was that, at the time, I was doing this mainly as a hobby from a computer in a shared venue to which I had occasional access.  The computer in question did not have Octave installed on it, and I did not want to have to pester the maintainers of it to install Octave.  I did not want to have to bother, at this stage, with moving files between the shared computer on which I was developing the software and my computers at home).  I think I still have a copy of this C code.

Hugo Sparks's method (which he claims is derived from a British Standard for clockmaking that he quotes) of approximating each wheel tooth addendum by a pair of arcs (sections of circles, that is) seemed to be more hassle than was worth, so instead I used the method that he mentions in passing ("What do real epicycloidal wheels look like?") of varying $\theta$ between 0 and 1 to generate the whole wheel addendum shape.

My C code essentially "fudged" the pinion addendum shape (NOT using a single arc as per Hugo Sparks's article, but an even cruder wedge-like shape: the assumption here was that the pinion addendum was not required to touch any part of the wheel, and so any shape was acceptable for the pinion addendum provided that it did not come into contact with the wheel).

Initially my C code just produced SVG pictures of the pinion and wheel shape.  I then enhanced it to also produce an input file for OpenSCAD.  I 3D-printed a pair (*intended* to mesh) of gears from this code (I still have these: 20 wheel teeth and 10 pinion teeth, if I rememeber rightly), but was very disappointed to discover that they did not mesh smoothly.

After looking at, and thinking about, these gears for a few days (if I remember rightly), I realised the reason for the problem: The construction of the wheel and pinion is *not* inherently asymmetric, as one might mistakenly believe from reading Hugo Sparks's article!  In a correctly-constructed wheel and pinion pair, each wheel addendum "rolls" against the pinion dedendum for part of the meshing "cycle", and each pinion addendum "rolls" against the wheel dedendum for part of the meshing "cycle".

Having the pinion addenda of a shape that does not touch the wheel dedenda at all is *not* an inherent feature of cycloidal gears: it is just an approximation of true cycloidal gearing that is presumably acceptable when the ratio:

   number of wheel teeth / number of pinion teeth is very high.  I imagine that, in that case, there is always more than one pinion dedendum in simultaneous contact with the wheel addenda, and so the lack of proper pinion addenda does not result in any "breaks" in the smooth contact between wheel and pinion: I have not attempted to analyse this properly).  Hence, I presume that this approximation of the pinion addendum shape is not acceptable for more moderate gearing ratios like 2 to 1.

Hence, I went "back to the drawing board" and attempted to implement my own true cycloidal gear generator in Octave, directly from the definitions of the hypocycloid and epicycloid.  If I remember rightly, I used the definitions in Wikipedia articles.

I prefer not to cite Wikipedia, so equivalent definitions are, if I am not mistaken, given in references [7] and [8].

Note that, at the time that I wrote my Octave cycloidal gear generator, I was unaware of the existence of a chapter (chapter 13) in Litvin's book (reference [4]) on cycloidal gearing.  Litvin's 1989 NASA technical report (reference [5]) *does not* appear to include any discussion of cycloidal gearing, and so I incorrectly assumed back in 2014 that Litvin's book also failed to discuss cycloidal gearing.  I was not aware of the chapter on cycloidal gearing in Litvin's book until about a fortnight ago (June 2016).

I can't be absolutely sure of the contents of Chapter 13 until I get access to the complete book, but I have seen the contents of

Litvin's book online, and also seen a selection of pages from Chapter 13 online.  It looks to me as if Chapter 13 includes similar definitions of the hypocycloid and epicycloid as are given in references [7] and [8], and proves that cycloidal gears mesh correctly (such as to obey the "fundamental law of gearing").  However, I'm pretty sure that it *does not* explicitly discuss the tooth tip problem or generating wheel step problem (we discussed these two problems in the abstract, and will discuss them in more detail below).

## **Advantages of Cycloidal over Involute Gearing:**

My initial personal reason for preferring cycloidal gearing to involute gearing, for my application, was, as I have said earlier, just the apparent conceptual simplicity of cycloidal gearing.

However, if Hugo Sparks is to be believed (I haven't had the inclination to try to verify these claims myself), cycloidal gearing has some clear advantages over involute gearing for clocks.  He says that the issue is "somewhat controversial", but "there seem to be three reasons":

> 1) *Manufacturing*: The pinions are easier to make. As we shall see, the acting part of the pinion leaf is a flat radial surface. Watch pinions are very tiny and to minimize friction, the acting surfaces must be polished. It's a lot easier to polish a flat surface without disturbing the geometry.
>
> 2) *Geometry*: The gear trains used in horology have multiple stages of wheels driving pinions where the pinions have very few teeth. Involute profiles for these pinions would be undercut, which would make them easy to break and difficult to manufacture.
>
> 3) *Friction*: Involute gears normally mesh with 2 or 3 teeth in contact at the same time while cycloidal gears can be designed so that only 1 or 2 teeth are in contact at a time. Because some sliding contact occurs in all gear designs, involute gearing will tend to have more friction that a comparable cycloidal design. In machinery where lubrication is used, the extra friction of involute gears is negligable and extra contact area is an advantage because it distributes the load. But it's desirable for watches and clocks to operate without lubrication on the gear teeth: They are expected to run for years without maintenance and, in clocks, in dusty unsealed cases. Lubricated teeth would attract dust and end up dirty and abrasive.

It would seem to me that reason (1) is also relevent to my application (heliostat gears rather than clock gears) if the pinion size is small, because of the relatively poor precision of

3D printers in comparison to ordinary CNC machines.  However, it might not be so important if prototyping of my proposed heliostat design is successful, resulting in the designs going into mass production with plastic moulding machines. (I presume that plastic moulding machines normally use CNC-machined moulds, although I am not very familiar with the plastics industry).

I certainly see reasons (2) and (3) as being relevent to my application: clearly heliostat gears, like clock gears (but *unlike* a car transmission) could be required to produce high torque (look at the size of an Ivanpah heliostat: however, I hope that I can get the cost of the gearing mechanism down enough that it would not be necessary to use heliostats as large as the Ivanpah ones), but transmit very little mechanical power.

I don't clearly remember where I put the file containing this calculation (I am happy to repeat the calculation if required), but I remember calculating in 2014 that a microwave oven fan motor, if geared down through about 15 stages of 2 to 1 gearing (as a first prototype: I expect that higher gearing ratios could be used to reduce the number of gears required, and hence reduce the production cost), can turn a heliostat through one revolution per hour (more than adequate, since we only need one revolution per day) with 500 Newton-metres of torque!  I remember being told at the time, by a visiting Australian man whose name I don't rememeber, that 500 Newton-metres is as much torque as a large Diesel truck (presumably at the wheels: I haven't researched this properly), i.e. I think more than enough even for an Ivanpah heliostat.

However, the amount of *mechanical power* delivered by the microwave oven fan motor is only, if I remember rightly (from the rated electrical power printed on the motor, which of course is an overestimate because of the losses in the motor, but I presume not much more than the output mechanical power of the motor at its most energy-efficient speed) about 30 watts.  If I remember righly, I assumed, for my calculations (I don't know enough about electrical engineering to be able to determine the speed that gives the maximum mechanical power output, and the corresponding maximum mechanical power output, properly), that the fan motor (a two-pole induction motor) can deliver 15 watts of mechanical power at *half* its synchronous speed.

## **Conceptual Design of my Octave Implementation:**

### Solving the Tooth Tip Problem

Hugo Sparks's iterative method of solving his equations (9) and (11) for two variables seemed overly complicated, and perhaps potentially succeptable to failure to converge.

Sorry, I have lost my original handwritten diagram of this from 2014, and I don't have the stamina to try to re-create it here, so I hope that you can visualise what is happening here. I am happy to add diagrams of this to this article if there is enough interest in this from readers.

Basically, we can assume that the gear (we presume it to be the wheel, although the same construction is used for the pinion) is centred on the origin, and rotated such that the bottom of a tooth addendum lies on the positive x axis. Then, from the number of teeth, it is trivial to calculate the angle, which we will call $\gamma$ , that the line through the origin and the tip of the tooth (i.e. the tip of the tooth addendum) must make with the positive x axis.

Clearly all tooth addenda have the same angular size; clearly also all tooth dedenda have the same angular size. These two angular sizes are the same: I have not bothered to prove that this last fact gives the correct geometry for the gears to mesh correctly, but I presume that this follows from the results in Litvin's Chapter 13. Certainly empirically, from the gear meshing animations that my Octave code can produce, it looks as if the wheel and pinion always mesh correctly. The wheel addendum generating circle (which is also the pinion dedendum generating circle; I presume that it is shown in Litvin's Chapter 13 that the pinion dedendum generating circle must be the same size as the wheel addendum generating circle for the wheel and pinion to mesh correctly) is of half the diameter of the pinion pitch circle so that the pinion dedenda are sections of radial straight lines (as with Hugo Sparks's construction of cycloidal gears). Similarly, the pinion addendum generating circle (which is also the wheel dedendum generating circle; again, I presume that Litvin's Chapter 13 shows that this is a requirement for the wheel and pinion to mesh correctly) is of half the diameter of the wheel pitch circle so that the wheel dedenda are sections of radial straight lines (again, as with Hugo Sparks's construction of cycloidal gears).

Hence, when the centre of the generating circle is on the x-axis, the point on the perimeter of the generating circle that generates the epicycloid touches the wheel's pitch circle. We'll call this point on the perimeter of the generating circle the "generating point" from now on.

The quantity that we *do not* know, however, is, the amount by which we have to roll the generating circle "up" the pitch circle to make the generating point "hit" the line through the origin which makes the angle $\gamma$ with the positive x axis (indicating that the generating point is at the tip of the tooth). More precisely, we do not know the angle, which we will call $\delta$ , that the line through the origin and *centre* of the generating circle makes with the x axis when the generating point is at the tip of the tooth.

It seemed to me that it would be straightforward to calculate this angle $\delta$ with Newton-Raphson iteration: I believe that Newton-Raphson iteration is a much more well-known and well-studied method than the apparently "ad-hoc" iteration used on Hugo Sparks's web page.

This is the method for solving the tooth tip problem that I successfully incorporated into my Octave code.

I remember struggling for a while in early 2014 to solve this problem: initially I was quite confused by the whole problem and failed to appreciate the difference between the angles $\gamma$ and $\delta$. My first attempt to solve this problem in my Octave code did not work at all.

Strangely, however, the solution suddenly became clear to me while I was in the audience (for my own interest) of a well-publicised public lecture on a topic almost totally unrelated to this problem. The lecture was the IPENZ Pickering Lecture (in Auckland) on space exploration by the director (who had presumably come to New Zealand specifically to give these lectures) of NASA's Jet Propulsion Laboratory, Dr Charles Elachi. I thank Dr Elachi and IPENZ for providing me with an interesting distraction from this problem; I think that this distraction somehow contributed to my ability to solve it!

I remember going back to the computer shortly (the next day, if I remember rightly) after Dr Elachi's lecture and successfully implementing the Newton-Raphson iteration for solving the tooth tip problem.

## Details of Implementation of Tooth Tip Problem Solution

I remember working out the details of the tooth tip problem solution (by hand) on paper in 2014, and then implementing them in Octave. Unfortunately, I have now misplaced these original handwritten notes, although I might eventually re-find them. However, I still have the Octave implementation, which seems to give accurate answers for "reasonable" inputs (reasonable numbers of wheel and pinion teeth), although I am not very confident of its ability to work for more extreme input values. Producing a robust, properly-documented implementation is an area for further work.

The basic idea is that the code tries to find the angle $\delta$ that makes the *gradient* of the line through the origin and the addendum tip (i.e. $\tan(\gamma)$ ) as close as possible to the *gradient* of the line through the origin and the generating point. I chose to work with gradients rather than angles at the time because I assumed (correctly, I still think) that this would simplify the algebra.

The function "gradient_difference" computes a quantity that is zero when $\delta$ reaches the correct value; hence our task (for both the wheel and pinion) is to find the value of $\delta$ that makes the value returned by "gradient_difference" as close as possible to zero. Unfortunately, I don't remember exacly what "gradient_difference" computes; I vaguely remember that it is not actually the difference between the two gradients, but a "cross-multiplied" version of it (in order to simpilfy the algebra) that has the same sign as the difference between the gradients, but a different absolute value.

I do remember, however, that (as you can see from the output below), the code first starts by stepping around the wheel and pinion pitch circles in one degree increments until the value of "gradient_difference" becomes non-negative. It then records the angles at which this happens.

It then uses the resulting angles as a starting point for Newton-Raphson iteration in order to find the exact (to the machine's precision, of course) angles that make "gradient_difference" equal to zero. These are the required values of $\delta$ for the wheel and pinion. The function "newton_raphson_step" does one step of the Newton-Raphson iteration.

As I said earlier, I am not very confident of the robustness of this implementation. Two obvious deficiencies in this implementation (although probably not the *only* potential sources of wrong results from this implementation) are:

1) I think that the method for finding the starting point for the iteration (stepping around the circles in one-degree increments) becomes unreliable if either the ratio of number of wheel teeth to number of pinion teeth, or the number of wheel teeth (and hence pinion teeth) itself, or possibly both, becomes large. This needs to be investigated properly. I remember having to fiddle around a bit with the implementation to get it to work on all of the inputs for which I tried it. One observation that I *did* make, which I think might be useful in attempting to improve the method for finding the optimal starting angle, is that (if I am not mistaken: I think it is geometrically obvious), *the pinion addendum sides tend toward involutes as the number of wheel teeth tends toward infinity with the number of pinion teeth held constant.*

*2)* At the moment, the number of steps in each of the two Newton-Raphson iterations is fixed at 20, and the code does not make any attempt to automatically check the accuracy of the result obtained at the end of the 20 iterations. It does, however, (as you can see from the output of the code below) print the angles calculated in iterations 16 to 20 so that the user can subjectively confirm that the Newton-Raphson iteration has converged. Clearly it would be possible to calculate the actual difference between the tooth tip angle $\gamma$ and the angle that the line through the origin and generating point makes with the x-axis, and stop the iteration when the absolute value of this difference gets below a certain value. However, clearly the appropriate stopping value will depend upon the machine's precision, and I am certainly not an expert in numerical analysis. I would be happy to have some advice on this from a numerical analyst!

## Solving the Generating Wheel Step Problem

We now know the starting and stopping angles for the centres of the generating circles (for both wheel and pinion). Hence, all we need to do in order to generate the wheel addenda is to roll the wheel generating circle between the starting and stopping angles, and trace the path of the generating point. (We then do the same with the pinion generating circle to generate the pinion addenda).

My original aim (now fulfilled to a certain extent: I have a version of the Octave code below that also outputs OpenSCAD code to 3D print the wheel and pinion, but unfortunately I don't remember where I put it) was to produce OpenSCAD code for 3D printing the wheel and pinion. OpenSCAD certainly does not (if I am not mistaken) support epicycloids as a shape, so they need to be approximated by some shape that is supported by OpenSCAD. Polygonal paths *are* supported by OpenSCAD (and presumably almost every other CAD program that is of any practical use), and so approximating the addendum sides by polygonal paths is an obvious solution.

I have never had the inclination to look into the details of the "slicing" algorithm (used to convert the low-level geometry output by OpenSCAD into the sequence of physical steps performed by the 3D printer's stepping motors) used for 3D printing to try to understand it. However, I presume that, if the minimum distance between any two adjacent points on a polygonal approximation to the true shape of an addendum is comfortably (I presume smaller than a fifth would be more than adequate) smaller than the 3D printer's advertised "resolution", then the distortion of the addendum shapes introduced by my polygonal approximation to epicycloids will be much smaller than the inherent mechanical limitations of the 3D printer's accuracy.

## The Problem

One obvious possibility would be to try "rolling" the generating circle (i.e. incrementing the generating circle angle) in steps of some fixed angle. However, if I am not mistaken (I believe that this is true, and I believe that it is very easy to find a counterexample to show this, so I can't be bothered to try to show it), this doesn't work properly: fixed-size generating circle increments do not give fixed distances between points on the output polygonal path.

Hence, we ideally need some method for calculating the (variable-sized) angle increments needed to give fixed distances between points on the output path.  Unfortunately, however, as far as I can see (from the equation defining an epicycloid given in, e.g. reference [7]), there is no analytical was to calculate these angle increments.

## Current Solution

The current solution, implemented in the function draw_tooth_half, is to divide the interval of angles that produces the whole path up into 20 equal-sized steps:

    angle_loop_increment = atr / 20;

However, as we remarked above, I do not believe that this gives equal distances between the points on the output path.  At the moment, I simply tolerate this, and assume that this gives good enough resolution for both SVG output images and 3D printing.  The output image, included in this document, produced with this method, looks good to me.  Clearly, however, this is not a good, robust solution.  One obvious problem with it is that if we make the gears very large, while keeping the assumed 3D printer resolution constant, the distance between output polygon points will become larger than the printer resolution.

Clearly having the number of steps "hard-coded" to 20 is highly undesirable: the constant "20" has nothing to do with any inherent property of the gears being generated.  I simply chose this value because it seemed reasonable for the size and resolution of gears that I was attempting to generate back in 2014, and appeared to give reasonably good results for that size and resolution.

## Proposed Priority Queue Algorithm

I puzzled over the problem of trying to automatically calculate satisfactory generating wheel step sizes for most of 2014, without finding any reasonable solution.  I had almost forgotton about this problem (I was busy with a job unrelated to generating cycloidal gears in 2015) until late 2015 (I don't remember the exact date) when I suddenly realised that there is a simple and, if it works as well as I anticipate, very robust solution, based on a priority queue.

Conceptually, the priority queue entries are triples representing segments of the polygonal output path:

$$(\theta_s, \theta_f, \delta)$$

where $\theta_s$ and $\theta_f$ are the generating circle angles corresponding to the start and finish of the segment, and $\delta$ is the length of the segment.  The priority queue is ordered by $\delta$ in *reverse* length, i.e. the *longest* segment (greatest $\delta$) has highest priority.

Then, in order to generate the path, we start by initialising the priority queue to be empty, and then putting a single segment into the priority queue, corresponding to a line from the tip of the dedendum (the end of the dedendum that touches the pitch circle) to the addendum tip.  For this segment:

a) $\theta_s$ is zero.
b) $\theta_f$ is the generating circle angle corresponding to the addendum tip, which we previously calculated by Newton-Raphson iteration (this is called *atr* in the function *draw_tooth_half*).
c) $\delta$ is calculated from $\theta_s$ and $\theta_f$.

To generate the complete path, we then keep on splitting (in half in generating-wheel-angle space, roughly speaking, but not exactly half in terms of Euclidian distance) the segment from the top of the priority queue, and then pushing the two parts of it back onto the priority queue, until the longest segment in the queue is at most a fixed length $\epsilon$.  (The length $\epsilon$ is the desired maximum distance between output points).

More precisely, the algorithm is as follows:

1) Pop the segment (the longest segment, by our above definitions) $(\theta_s, \theta_f, \delta)$ from the top of the priority queue.
2) If $\delta \leq \epsilon$ then we are finished, so halt.
3) Let $\theta_h$ be the angle halfway between $\theta_s$ and $\theta_f$, i.e.:
$$\frac{\theta_s + \theta_f}{2}$$
4) Create the two segments $(\theta_s, \theta_h, \delta_1)$ and $(\theta_h, \theta_f, \delta_2)$, with the lengths $\delta_1$ and $\delta_2$ calculated from the generating point coordinates corresponding to the three generating circle angles $\theta_s$, $\theta_h$ and $\theta_f$.
5) Push the two segments $(\theta_s, \theta_h, \delta_1)$ and $(\theta_h, \theta_f, \delta_2)$ onto the priority queue.
6) Go back to step 1.

Although this algorithm does not preduce a polygonal path of segments of length *exactly* $\epsilon$, I believe that it comes reasonably close:

A) Certainly all segments are of length *at most* $\epsilon$.
B) I don't think that it can produce any output segments of length less than $\epsilon/2$, although I have yet to test the algorithm (I certainly do not have any proof of this).

However, it appears that Octave does not have a built-in priority queue implementation, nor any convenient means of implementing a priority queue.  Hence, my intention is to re-write the whole implementation in another language which has better support for priority queues.  I am considering the following two options:

1) A functional language, probably Haskell (Haskell is similar to Gofer; I was taught Gofer as a student in 1995, but it appears that Gofer is no longer being maintained: see reference [10]).  Thanks to Dr Radu Nicolescu of the Computer Science Department, University of Auckland, for advice on functional programming.

2) Java, which appears to have (in the standard class library) an implementation of a priority queue.

## Proposed Newton-Raphson Stepping Implementation

About a fortnight ago, while I was working on this document, I thought of another approach to solving the generating wheel step problem which, if I am not mistaken, could produce a polygonal approximation to the epicycloid with equal distances between the points.  The idea is simple in principle:

1) As with the priority queue solution, we calculate, based upon the mechanics of the 3D printer, the desired distance (here exact, not an upper bound) $\epsilon$ between the desired points on the output polygon.

2) Let $\theta_1$ be zero.

3) Let *i*, the loop iteration counter, be one.

4) Let $\theta_f$ be the generating circle angle corresponding to the addendum tip, as with the priority queue implementation above.

5) If the distance between the generating point positions corresponding to the angles $\theta_i$ and $\theta_f$ is less than or equal to $\epsilon$ then halt (we are within distance $\epsilon$ of the end of the epicycloid).

6) Find, by Newton-Raphson iteration, the angle $\theta_{i+1}$ for which the distance between the generating point positions corresponding to angles $\theta_i$ and $\theta_{i+1}$ is *exactly* (to the machine's precision, of course) equal to $\epsilon$ .

7) Increment *i*.

8) Go back to step 5.

I have not yet tried to implement this algorithm, however.  One issue that would have to be resolved in order to implement it correctly is finding a good starting point for the calculation of $\theta_{i+1}$ in step (5). If $\theta_i$ always works, then great: the problem is solved! (*Proving* that $\theta_i$ always works, i.e. always causes the iteration to converge, might not be easy, however).

If starting the Newton-Raphson iteration from $\theta_i$ does not reliably work, then another possibly-viable approach would be to use the priority queue algorithm of the previous section to find an *initial* approximation to the polygonal approximation to the epicycloid, with the $\epsilon$ value of the priority queue algorithm being considerably smaller (maybe a tenth or a twentieth) of the $\epsilon$ value input to *this* algorithm. The output path could then be tranferred to from the priorty queue to a balanced search tree that also supports sequential search (e.g. B+-tree), ordered by generating-circle angle. To find the starting point for a given input angle $\theta_i$ , we use the algorithm:

1) Find the angle $\varphi_i$ closest to $\theta_i$ in the search tree.

2) Step through the search tree elements from $\varphi_i$ until you hit the first search tree element $\varphi_j$ for which the distance between the generating point positions corresponding to angles $\theta_i$ and $\varphi_j$ is greater than or equal to $\epsilon$ (the $\epsilon$ of *this* algorithm, that is, not the $\epsilon$ of the priority queue algorithm).

3) The angle $\varphi_j$ is the starting point for the Newton-Raphson iteration.

(In fact, using a tree might not be necessary; it might be possible, although maybe less elegant, simply to put the priority queue entries into a linked list, ordered by generating-circle angle, and then step through it sequentially, saving the position in the linked list after each iteration on the "outer" loop as the starting point for the next iteration of the "outer" loop. However, a singly-linked list might not be adequate, since it might be necessary to step back and forwards through the list to find the correct angle $\varphi_i$ ; clearly this needs to be investigated further).

**Note that I have neither a proof, nor any emperical evidence, that the proposed Newton-Raphson iteration here will converge. This clearly needs to be investigated.**

## **Source Code of my Octave implementation**

```
#!/usr/bin/octave

                # Octave cycloidal gear generator

global wheel_pitch_circle_radius;
global pinion_pitch_circle_radius;
```

```
global wheel_dedendum_depth;
global pinion_dedendum_depth;
global optimal_wheel_addendum_tip_radians;
global optimal_pinion_addendum_tip_radians;
global wheel_radius_ratio;
global pinion_radius_ratio;
global wheel_tooth_tip_angle;
global pinion_tooth_tip_angle;
global scale_factor;
global current_x;
global current_y;
global leftmost_x;
global bottommost_y;
global svg_height;
global wtp_xshift;
global wtp_yshift;
global clip_dedenda;

scale_factor = 15; # initially undefined, calculated later
current_x = nan; # unscaled
current_y = nan; # unscaled
svg_height = 10;
svg_width = 10;

                    # WTP stands for "wheel to pinion"

#disp(argv);

wheel_teeth = input("Number of wheel teeth ->");   # Input parameter
pinion_teeth = input("Number of pinion teeth ->"); # ditto
printf("Simulate meshing? (0 for no, 1 for single mesh, 2 for animation)?\n");
simulate_mesh = input("Simulate meshing? ->");
if ((simulate_mesh != 0) && (simulate_mesh != 1) && (simulate_mesh != 2))
  printf("Invalid value given for simulate_meshing.\n");
  exit(1)
endif

printf("Clip dedenda to inner circle? (0 for no, 1 for yes)?\n");
clip_dedenda = input("Clip dedenda? ->");
if ((clip_dedenda != 0) && (clip_dedenda != 1))
   printf("Invalid value given for clip_dedenda.\n");
   exit(1);
endif

svg_height = input("SVG output height (default 500) ->");
svg_width = input("SVG output width (default 500) ->");
if (simulate_mesh == 0)
  printf("Will not do meshing simulation.\n");
  no_mesh = 1;
  single_mesh = 0;
  mesh_animation = 0;
elseif (simulate_mesh == 1)
  printf("Will do single mesh simulation.\n");
  no_mesh = 0;
  single_mesh = 1;
  mesh_animation = 0;
elseif (simulate_mesh == 2)
  printf("Will do meshing simualtion animation.\n");
  num_frames = input("Number of frames ->");
  no_mesh = 0;
  single_mesh = 0;
  mesh_animation = 1;
```

```
else
  printf("Invalid value of meshing simulation chosen.\n");
  exit(1);
endif

if (pinion_teeth > wheel_teeth)
  printf("pinion cannot have more teeth than wheel\n");
  exit(1);
endif

                    # module is pitch circle diamater over
                    # number of teeth, IIRR

                    # Ratio of wheel and pinion pitch circle
                    # circumferences (and hence radii)
                    # equals ration of wheel and pinion
                    # numbers of teeth, and hence wheel and
                    # pinion have the same module.

function terminate_polygon
  global current_x;
  global current_y;

  current_x = nan;
  current_y = nan;
endfunction

function dump_angle(angle_radians)
     angle_degrees = angle_radians * (360 / (2 * pi));
     printf("    Angle in degrees: %5.5f\n",angle_degrees);
     printf("    Angle in radians: %5.5f\n",angle_radians);
endfunction

wheel_pitch_circle_diameter = wheel_teeth; # in modules
wheel_pitch_circle_radius = wheel_teeth / 2;

pinion_pitch_circle_diameter = pinion_teeth; # in modules
pinion_pitch_circle_radius = pinion_teeth / 2;

one_over_root_two = 1/sqrt(2);

if (no_mesh)
  wtp_xshift = pinion_pitch_circle_radius + wheel_pitch_circle_radius;
  wtp_yshift = pinion_pitch_circle_radius + wheel_pitch_circle_radius;
endif

if (single_mesh || mesh_animation)
  radii_sum = pinion_pitch_circle_radius + wheel_pitch_circle_radius;
  wtp_xshift = one_over_root_two * radii_sum;
  wtp_yshift = one_over_root_two * radii_sum;
endif

printf("Wheel teeth: %d, Pinion teeth: %d\n",wheel_teeth,pinion_teeth);
printf("Wheel PCR: %5.5f\n",wheel_pitch_circle_radius);
printf("Pinion PCR: %5.5f\n\n",pinion_pitch_circle_radius);

wheel_generating_circle_radius = pinion_pitch_circle_radius / 2;
pinion_generating_circle_radius = wheel_pitch_circle_radius / 2;

printf("Wheel GCR: %5.5f\n",wheel_generating_circle_radius);
printf("Pinion GCR: %5.5f\n\n",pinion_generating_circle_radius);
```

```
wheel_radius_ratio = wheel_generating_circle_radius / wheel_pitch_circle_radius;
wheel_gc_one_rev_degrees = 360*wheel_radius_ratio;

printf("Wheel GC one rev degrees: %5.5f\n",wheel_gc_one_rev_degrees);

pgcr = pinion_generating_circle_radius;
ppcr = pinion_pitch_circle_radius;

pinion_radius_ratio = pgcr / ppcr;

pinion_gc_one_rev_degrees = 360*pinion_radius_ratio;

printf("Pinion GC one rev degrees: %5.5f\n",pinion_gc_one_rev_degrees);

if (pinion_gc_one_rev_degrees > 360)
  printf("pinion GC one rev degrees > 360, so cutting back to 360\n");
  printf("(Clearly optimum pinion GC angle should be within one rev\n");
  pinion_gc_one_rev_degrees = 360;
endif

printf("Wheel RR: %5.5f\n",wheel_radius_ratio);
printf("Pinion RR: %5.5f\n\n",pinion_radius_ratio);

wheel_tooth_tip_angle = 2*pi/(wheel_teeth*4);

pinion_tooth_tip_angle = 2*pi/(pinion_teeth*4);

printf("Wheel tooth tip angle:\n");
dump_angle(wheel_tooth_tip_angle);

printf("Pinion tooth tip angle\n");
dump_angle(pinion_tooth_tip_angle);

wheel_tooth_tip_gradient = tan(wheel_tooth_tip_angle);
pinion_tooth_tip_gradient = tan(pinion_tooth_tip_angle);

printf("Tooth tip gradiants:\n");
printf("Wheel TTG: %5.5f\n",wheel_tooth_tip_gradient);
printf("Pinion TTG: %5.5f\n\n",pinion_tooth_tip_gradient);

if (wheel_tooth_tip_gradient > 1)
      printf("Wheel tooth gradient > 1: that's unreasonable.\n");
      exit(1);
endif

function ret_value = gradient_difference(r,m,theta)
  term_1 = (1 + r) * sin(theta);
  term_2 = -r * sin (theta + theta / r);
  term_3 = -(m + m*r) * cos (theta);
  term_4 = m * r * cos (theta + theta / r);

  ret_value = term_1 + term_2 + term_3 + term_4;
endfunction

function ret_value = newton_raphson_step(r,m,theta)
  numerator = gradient_difference(r,m,theta);

  dem_term_1 = (1 + r) * cos(theta);
  dem_term_2 = -r * cos(theta + theta/r) - cos(theta + theta/r);
  dem_term_3 = (m + m*r) * sin(theta);
  dem_term_4 = - m * r * sin(theta + theta/r) - m*sin(theta + theta/r);
```

```
  denominator = dem_term_1 + dem_term_2 + dem_term_3 + dem_term_4;

  ret_value = theta - numerator / denominator;
endfunction

function ret_value = addendum_tip_height(pcr,r,theta)
      # distance from gear centre to point on addendum tip
      # corresponding to given generating circle angle.
  x = cos(theta) + r*cos(theta) - r * cos(theta + theta/r);
  y = sin(theta) + r*sin(theta) - r * sin(theta + theta/r);
  ret_value = pcr*sqrt(x*x + y*y);
endfunction

function ret_value = wheel_tip_height(r,theta)
  global wheel_pitch_circle_radius;

  ret_value = addendum_tip_height(wheel_pitch_circle_radius,r,theta);
endfunction

function ret_value = pinion_tip_height(r,theta)
  global pinion_pitch_circle_radius;

  ret_value = addendum_tip_height(pinion_pitch_circle_radius,r,theta);
endfunction

function ret_value = rotated_addendum_tip(pcr,r,theta,rot_angle,flip,rot2)
                      # rot_angle is angle (+ve for
                      # anticlockwise, -ve for clockwise)
                      # through which resulting gear should be
                      # rotated.

                             # if flip is 1, result of first
                             # rotation is flipped about x axis
                             # and then rotated through angle rot2
                             # (rot2 uses same sign convention as
                             # rot_angle).

                             # if flip is 0, neither flipping nor
                             # second rotation is performed.
  phi = theta / r;
  x = cos(theta) + r * cos(theta) - r * cos(theta + phi);
  y = sin(theta) + r * sin(theta) - r * sin(theta + phi);

  flip_matrix = [1, 0; 0, -1];
  rot1_matrix = [cos(rot_angle), -sin(rot_angle);
             sin(rot_angle), cos(rot_angle)];
  rot2_matrix = [cos(rot2), -sin(rot2); sin(rot2), cos(rot2)];

  rot1_result = rot1_matrix * [x; y];

  if (flip)
    flip_result = flip_matrix * rot1_result;
    ret_value = pcr * rot2_matrix * flip_result;
  else
    ret_value = pcr * rot2_matrix * rot1_result;
  endif
endfunction

function ret_value = rad_to_deg(input_angle)
  ret_value = 360 * (input_angle/(2*pi));
endfunction
```

```
function ret_value = deg_to_rad(input_angle)
   ret_value = 2*pi * (input_angle / 360);
endfunction

function return_value = svg_x(xinput)
   global scale_factor;
   global svg_centre_x;
   global leftmost_x;

   shifted_input = xinput - leftmost_x;
   return_value = floor(scale_factor * shifted_input);
%  printf("svg_x called: output number = %d\n",return_value);
   if (!finite(return_value))
       printf("Error: svg_x outputting infinite value: halting\n");
       exit(1);
   endif
endfunction

function return_value = svg_y(yinput)
   global scale_factor;
   global svg_centre_y;
   global bottommost_y;
   global svg_height;

   shifted_input = yinput - bottommost_y;
   return_value = svg_height - floor(scale_factor * shifted_input);
%  printf("svg_y called: output number = %d\n",return_value);
   if (!finite(return_value))
       printf("svg_y outputting infinite value: halting\n");
       exit(1);
   endif
endfunction

function clipped_coords = clip_to_radius(input_x,input_y,clip_radius,
                                centre_x,centre_y,outer_x,outer_y)
# the purpose of outer_x and outer_y is to specify the direction vector of the
# line from the centre, particularly when (input_x,input_y) is the origin:
# basically, (outer_x,outer_y) is the end of the line that is not
(centre_x,centre_y).

   x_displacement = input_x - centre_x;
   y_displacement = input_y - centre_y;
   outer_x_displacement = outer_x - centre_x;
   outer_y_displacement = outer_y - centre_y;

%  printf("clip_to_radius called\n");
%  printf("  input_x = %5.5f, ",input_x);
%  printf("  input_y = %5.5f, ",input_y);
%  printf("  outer_x = %5.5f, ",outer_x);
%  printf("  outer_y = %5.5f, ",outer_y);
%  printf("  clip_radius = %5.5f\n",clip_radius);
%  printf("  centre_x = %5.5f, centre_y = %5.5f\n",centre_x,centre_y);

   outer_radius = sqrt((outer_x_displacement)^2 + (outer_y_displacement)^2);
   input_radius = sqrt((x_displacement)^2 + (y_displacement)^2);
%  printf("  outer_radius = %5.5f\n",outer_radius);
%  printf("  input_radius = %5.5f\n",input_radius);

   if (input_radius < clip_radius)
       normalized_x_displacement = outer_x_displacement / outer_radius;
       normalized_y_displacement = outer_y_displacement / outer_radius;
```

```
      clipped_x_displacement = normalized_x_displacement * clip_radius;
      clipped_y_displacement = normalized_y_displacement * clip_radius;
   else
      clipped_x_displacement = x_displacement;
      clipped_y_displacement = y_displacement;
   endif

   clipped_coords = [clipped_x_displacement + centre_x,
                     clipped_y_displacement + centre_y];
%  printf("   clipped_coords = (%5.5f, %5.5f)\n",clipped_coords(1),clipped_coords(2));
endfunction

function \
line_to(output_file,graph_x,graph_y,r,g,b,centre_x,centre_y,base_radius)
   global current_x;
   global current_y;
   global clip_dedenda;

   if (isnan(current_x))
                         # Start new polygon
      current_x = graph_x;
      current_y = graph_y;
   else
      start_x = current_x;
      start_y = current_y;
      finish_x = graph_x;
      finish_y = graph_y;
      draw_line = 0;
%     printf("line_to called: centre_x = %5.5f, centre_y = %5.5f\n",centre_x,centre_y);
%     printf("   current_x = %5.5f, current_y = %5.5f\n",current_x,current_y);
%     printf("   graph_x = %5.5f, graph_y = %5.5f, base_radius = %5.5f\n",graph_x,graph_y,base_radius);

      start_radius = sqrt((start_x - centre_x)^2 + (start_y - centre_y)^2);
      finish_radius = sqrt((finish_x - centre_x)^2 + (finish_y - centre_y)^2);

      if (clip_dedenda)
         if (start_radius < base_radius && finish_radius >= base_radius)
            clipped_coords = clip_to_radius(start_x,start_y,base_radius,
                                  centre_x,centre_y,finish_x,finish_y);
            start_x = clipped_coords(1);
            start_y = clipped_coords(2);
            draw_line = 1;
         endif
         if (finish_radius < base_radius && start_radius >= base_radius)
            clipped_coords = clip_to_radius(finish_x,finish_y,base_radius,
                                  centre_x,centre_y,start_x,start_y);
            finish_x = clipped_coords(1);
            finish_y = clipped_coords(2);
            draw_line = 1;
         endif
         if (start_radius < base_radius && finish_radius < base_radius)
            draw_line = 0;
         endif
         if (start_radius >= base_radius && finish_radius >= base_radius)
            draw_line = 1;
         endif
      else
         draw_line = 1;
```

```
      endif

      if (draw_line)
      fprintf(output_file,"<line x1=\"%d\"",floor(svg_x(start_x)));
      fprintf(output_file," y1=\"%d\"",floor(svg_y(start_y)));
      fprintf(output_file," x2=\"%d\"",floor(svg_x(finish_x)));
      fprintf(output_file," y2=\"%d\"",floor(svg_y(finish_y)));
      fprintf(output_file," style=\"stroke:rgb(%d,%d,%d);",r,g,b);
      fprintf(output_file,"stroke-width:2\" />\n");
      endif
    current_x = graph_x;
    current_y = graph_y;
  endif
endfunction

function back_to_centre(output_file,xshift,yshift,r,g,b,clip_radius)
  line_to(output_file,xshift,yshift,r,g,b,xshift,yshift,clip_radius);
endfunction

function \
draw_tooth_half(output_file,atr,rr,tta,pcr,flip,rot2,xshift,clip_radius)
  global current_x;
  global current_y;
                      # optimal_wheel_addendum_tip_radians now
                      # atr

                      #  global wheel_radius_ratio now rr

                      #  global wheel_tooth_tip_angle now tta
                      #  global wheel_pitch_circle_radius now pcr

  angle_loop = 0;
  angle_loop_increment = atr / 20;
                      # FIXME: step size should be changable

  while (angle_loop <= atr)
    r = rr;
    if (flip)
      theta = atr - angle_loop;
    else
      theta = angle_loop;
    endif # reverses direction of loop so there is no "jump" in the
                      # graph between the two halves when
                      # draw_tooth_half is called for both
                      # tooth halves.

    rot1 = -1*tta;

    wheel_tip_point = rotated_addendum_tip(pcr,r,theta,rot1,flip,rot2);
    graph_x = wheel_tip_point(1) + xshift;
    graph_y = wheel_tip_point(2) + xshift;

    line_to(output_file,graph_x,graph_y,0,0,0,xshift,xshift,clip_radius);

    angle_loop = angle_loop + angle_loop_increment;
  endwhile
endfunction

function draw_circle(output_file,circle_radius,xshift,yshift,colour)
  global scale_factor;

  svg_radius = floor(scale_factor * circle_radius);
```

```
   svg_centre_x = svg_x(xshift);
   svg_centre_y = svg_y(yshift);

   fprintf(output_file,"<circle cx=\"%d\" ",svg_centre_x);
   fprintf(output_file,"cy=\"%d\" ",svg_centre_y);
   fprintf(output_file,"r=\"%d\" ",svg_radius);
   fprintf(output_file,"stroke=\"%s\" ",colour);
   fprintf(output_file,"stroke-width=\"3\" style=\"fill:none\" />\n");
endfunction

optimal_wheel_addendum_tip_radians = 0;
angle_degrees = 0;
stop_loop = 0;

#printf("Difference between wheel tooth tip gradient and its optimal value\n");

printf("Finding wheel GC angle\n");

while (!stop_loop)
  angle_radians = (angle_degrees / 360.0) * 2 * pi;

  r = wheel_radius_ratio;
  m = wheel_tooth_tip_gradient;
  grad_diff = gradient_difference(r,m,angle_radians);
  tip_height = wheel_tip_height(r,angle_radians);

  printf("angle in degrees: %5.5f, grad_diff: %5.5f\n",angle_degrees,grad_diff);

  if (grad_diff > 0)
    optimal_wheel_addendum_tip_radians = angle_radians;
    stop_loop = 1;
  endif

  if (angle_degrees > 360.0)
    printf("error: could not find wheel GC stopping angle\n");
    exit(1);
  endif

  angle_degrees = angle_degrees + 1.0;
endwhile

optimal_pinion_addendum_tip_radians = 0;
angle_degrees = 0;
stop_loop = 0;

printf("Finding pinion GC angle:\n");

while(!stop_loop)
  angle_radians = (angle_degrees / 360.0) * 2 * pi;

  r = pinion_radius_ratio;
  m = pinion_tooth_tip_gradient;
  grad_diff = gradient_difference(r,m,angle_radians);
  tip_height = pinion_tip_height(r,angle_radians);

  printf("angle in degrees: %5.5f, grad_diff: %5.5f\n",angle_degrees,grad_diff);

  if (grad_diff > 0)
    optimal_pinion_addendum_tip_radians = angle_radians;
    stop_loop = 1;
```

```
    endif

    if (angle_degrees > 360.0)
      printf("error: could not find pinion GC stopping angle\n");
      exit(1);
    endif

    angle_degrees = angle_degrees + 1.0;
endwhile

printf("First approximation of wheel GC angle for addendum tip\n");
dump_angle(optimal_wheel_addendum_tip_radians);

printf("First approximation of pinion GC angle for addendum tip\n");
dump_angle(optimal_pinion_addendum_tip_radians);

printf("\n\nNow starting Newton-Raphson iteration for wheel GC\n\n");

old_angle = optimal_wheel_addendum_tip_radians;

for newton_loop=(0:20)
  r = wheel_radius_ratio;
  m = wheel_tooth_tip_gradient;

  new_angle = newton_raphson_step(r,m,old_angle);
  if (newton_loop > 15)
    printf("After iteration step %d, angle is:\n",newton_loop);
    dump_angle(new_angle);
  endif
  old_angle = new_angle;
endfor

optimal_wheel_addendum_tip_radians = new_angle;

printf("\nNow starting NR iteration for pinion GC\n\n");

old_angle = optimal_pinion_addendum_tip_radians;

for newton_loop=(0:20)
  r = pinion_radius_ratio;
  m = pinion_tooth_tip_gradient;

  new_angle = newton_raphson_step(r,m,old_angle);
  if (newton_loop > 15)
    printf("After iteration step %d, angle is:\n",newton_loop);
    dump_angle(new_angle);
  endif
  old_angle = new_angle;
endfor

optimal_pinion_addendum_tip_radians = new_angle;

if (no_mesh)
  wp_file = fopen ("wheel_pinion_nomesh.svg", "w");
endif

if (single_mesh)
  wp_file = fopen ("wheel_pinion_single_mesh.svg", "w");
endif

if (single_mesh || no_mesh)
  fprintf(wp_file,"<svg height=\"%d\"",svg_height);
```

```
    fprintf(wp_file,"   width=\"%d\">\n",svg_width);

    if (wp_file < 0)
      printf("Could not create wheel file.\n");
      exit(1);
    endif
  endif

  r = wheel_radius_ratio;
  theta = optimal_wheel_addendum_tip_radians;

  wheel_addendum_tip_height = wheel_tip_height(r,theta);

  printf("Wheel addendum tip height: %5.5f\n",wheel_addendum_tip_height);

  wheel_bound_radius = wheel_addendum_tip_height + 1.0;

  r = pinion_radius_ratio;
  theta = optimal_pinion_addendum_tip_radians;

  pinion_addendum_tip_height = pinion_tip_height(r, theta);

  printf("Pinion addendum tip height: \
          %5.5f\n",pinion_addendum_tip_height);

  pinion_bound_radius = pinion_addendum_tip_height + 1.0;

  printf("Wheel bound radius: %5.5f\n",wheel_bound_radius);
  printf("Pinion bound radius: %5.5f\n",pinion_bound_radius);

  leftmost_x = -wheel_bound_radius;
  topmost_y = pinion_bound_radius + wtp_yshift;

  rightmost_x = pinion_bound_radius + wtp_xshift;
  bottommost_y = -wheel_bound_radius;

  printf("leftmost x: %5.5f\n",leftmost_x);
  printf("bottommost y: %5.5f\n",bottommost_y);
  printf("rightmost x: %5.5f\n",rightmost_x);
  printf("topmost y: %5.5f\n",topmost_y);

  unscaled_width = rightmost_x - leftmost_x;
  unscaled_height = topmost_y - bottommost_y;

  printf("unscaled height: %5.5f\n",unscaled_height);
  printf("unscaled width: %5.5f\n",unscaled_width);

  printf("svg height: %d\n",svg_height);
  printf("svg width: %d\n",svg_width);

  vert_scale_factor = svg_height / unscaled_height;
  horiz_scale_factor = svg_width / unscaled_width;
  scale_factor = min(vert_scale_factor,horiz_scale_factor);

  printf("vertical scale factor: %5.5f\n",vert_scale_factor);
  printf("horiz scale factor: %5.5f\n",horiz_scale_factor);
  printf("smaller scale factor: %5.5f\n",scale_factor);

                        # N.B.: this INCLUDES pitch circle radius

  wheel_dedendum_depth = pinion_addendum_tip_height - pinion_pitch_circle_radius;
  pinion_dedendum_depth = wheel_addendum_tip_height - wheel_pitch_circle_radius;
```

```
  wheel_tta_degrees = rad_to_deg(wheel_tooth_tip_angle);
  pinion_tta_degrees = rad_to_deg(pinion_tooth_tip_angle);

function \
      draw_wheel_teeth(wheel_file,wheel_teeth,overall_wheel_rotation,pcr)
  global optimal_wheel_addendum_tip_radians;
  global wheel_radius_ratio;
  global wheel_tooth_tip_angle;
  global wheel_pitch_circle_radius;
  global pinion_pitch_circle_radius;
  global wheel_dedendum_depth;
  global pinion_dedendum_depth;

  wheel_base_radius = wheel_pitch_circle_radius - wheel_dedendum_depth;
  pinion_base_radius = pinion_pitch_circle_radius - pinion_dedendum_depth;

  wbr_with_clearance = wheel_base_radius - 0.5;

  pbr_with_clearance = pinion_base_radius - 0.5;

  for tooth_loop = (0:(wheel_teeth - 1))
    angle_between_teeth = 360 / wheel_teeth;
    tooth_angle = angle_between_teeth * tooth_loop + overall_wheel_rotation;
    tooth_angle_radians = deg_to_rad(tooth_angle);

    back_to_centre(wheel_file,0,0,0,0,0,wbr_with_clearance);

    atr = optimal_wheel_addendum_tip_radians;
    rr = wheel_radius_ratio;
    tta = wheel_tooth_tip_angle;

    draw_tooth_half(wheel_file,atr,rr,tta,pcr,0,
              tooth_angle_radians,0,wbr_with_clearance);
    draw_tooth_half(wheel_file,atr,rr,tta,pcr,1,
              tooth_angle_radians,0,wbr_with_clearance);

    back_to_centre(wheel_file,0,0,0,0,0,wbr_with_clearance);
  endfor

  terminate_polygon;
endfunction

function \
      draw_pinion_teeth(pinion_file,pinion_teeth,overall_pinion_rotation,pcr)
  global optimal_pinion_addendum_tip_radians;
  global pinion_radius_ratio;
  global pinion_tooth_tip_angle;
  global wtp_xshift;
  global wtp_yshift;
  global wheel_pitch_circle_radius;
  global pinion_pitch_circle_radius;
  global wheel_dedendum_depth;
  global pinion_dedendum_depth;

  wheel_base_radius = wheel_pitch_circle_radius - wheel_dedendum_depth;
  pinion_base_radius = pinion_pitch_circle_radius - pinion_dedendum_depth;

  wbr_with_clearance = wheel_base_radius - 0.5;

  pbr_with_clearance = pinion_base_radius - 0.5;
```

```
    for tooth_loop = (0:(pinion_teeth - 1))
      angle_between_teeth = 360 / pinion_teeth;
      tooth_angle = angle_between_teeth * tooth_loop + overall_pinion_rotation;
      tooth_angle_radians = deg_to_rad(tooth_angle);

      back_to_centre(pinion_file,wtp_xshift,wtp_yshift,0,0,0,pbr_with_clearance);

      atr = optimal_pinion_addendum_tip_radians;
      rr = pinion_radius_ratio;
      tta = pinion_tooth_tip_angle;

      draw_tooth_half(pinion_file,atr,rr,tta,pcr,0,tooth_angle_radians,
              wtp_xshift,pbr_with_clearance);
      draw_tooth_half(pinion_file,atr,rr,tta,pcr,1,tooth_angle_radians,
              wtp_xshift,pbr_with_clearance);

      back_to_centre(pinion_file,wtp_xshift,
              wtp_yshift,0,0,0,pbr_with_clearance);
    endfor

    terminate_polygon;
  endfunction

  if (no_mesh)
    overall_wheel_rotation = 0;
    overall_pinion_rotation = 0;

    pcr = wheel_pitch_circle_radius;
    draw_wheel_teeth(wp_file,wheel_teeth,overall_wheel_rotation,pcr);
    pcr = pinion_pitch_circle_radius;
    draw_pinion_teeth(wp_file,pinion_teeth,overall_pinion_rotation,pcr);
  endif

  if (single_mesh)
    pinion_tooth_period_angle = 360 / pinion_teeth;
    pinion_angle_increment = pinion_tooth_period_angle / 2;

    overall_wheel_rotation = 45;
    overall_pinion_rotation = 45 + pinion_angle_increment;
                    # so that wheel and pinion mesh
                    # correctly, regardless of number of
                    # teeth.

    pcr = wheel_pitch_circle_radius;
    draw_wheel_teeth(wp_file,wheel_teeth,overall_wheel_rotation,pcr);
    pcr = pinion_pitch_circle_radius;
    draw_pinion_teeth(wp_file,pinion_teeth,overall_pinion_rotation,pcr);
  endif

  function draw_wheel_pinion_circles(output_file)
    global wheel_pitch_circle_radius;
    global pinion_pitch_circle_radius;
    global wheel_dedendum_depth;
    global pinion_dedendum_depth;
    global wtp_xshift;
    global wtp_yshift;

    wheel_base_radius = wheel_pitch_circle_radius - wheel_dedendum_depth;
    pinion_base_radius = pinion_pitch_circle_radius - pinion_dedendum_depth;

    wbr_with_clearance = wheel_base_radius - 0.5;
```

```
    pbr_with_clearance = pinion_base_radius - 0.5;

    draw_circle(output_file,wheel_base_radius,0,0,"red");
    terminate_polygon;
    draw_circle(output_file,wbr_with_clearance,0,0,"blue")
    terminate_polygon;
    draw_circle(output_file,pinion_base_radius,wtp_xshift,wtp_yshift,"red");
    terminate_polygon;
    draw_circle(output_file,pbr_with_clearance,wtp_xshift,wtp_yshift,"blue");
    terminate_polygon;
endfunction

if (no_mesh || single_mesh)
  draw_wheel_pinion_circles(wp_file);
endif

if (mesh_animation)
                        # first attemot: wheel rotated
                        # anticlockwise (right) while pinion
                        # also rotated anticlockwise (wrong).

  wheel_tooth_period = 360.0 / wheel_teeth;
  pinion_tooth_period = 360.0 / pinion_teeth;

  wheel_frame_rotation = wheel_tooth_period / num_frames;
  pinion_frame_rotation = pinion_tooth_period / num_frames;
                        # last frame should be one frame short
                        # of being identical to first frame, so
                        # that animation flows smoothly if
                        # looped endlessly.

  converter_script = fopen("convert_script","w");
  fprintf(converter_script,"#!/bin/bash\n");
  frame_log = fopen("frame_log.txt","w");

  fprintf(frame_log,"wheel teeth: %d\n",wheel_teeth);
  fprintf(frame_log,"pinion teeth: %d\n",pinion_teeth);
  fprintf(frame_log,"svg width: %d\n",svg_width);
  fprintf(frame_log,"svg height: %d\n",svg_height);
  fprintf(frame_log,"num_frames: %d\n",num_frames);
  fprintf(frame_log,"wheel tooth period: ");
  fprintf(frame_log,"%5.5f degrees\n",wheel_tooth_period);
  fprintf(frame_log,"pinion tooth period: ");
  fprintf(frame_log,"%5.5f degrees\n",pinion_tooth_period);

  for frame_loop = (1:num_frames)
    printf("producing frame %d of %d\n",frame_loop,num_frames);
    frame_file_name = sprintf("frame%05d.svg",frame_loop);
    frame_file = fopen (frame_file_name, "w");

    fprintf(converter_script,
      "echo converting frame %d from svg to png\n",frame_loop);
    fprintf(converter_script,"convert ");
    fprintf(converter_script,"frame%05d.svg ",frame_loop);
    fprintf(converter_script,"frame%05d.png\n",frame_loop);

    fprintf(frame_file,"<svg height=\"%d\"",svg_height);
    fprintf(frame_file,"  width=\"%d\">\n",svg_width);

    draw_wheel_pinion_circles(frame_file);

    pinion_angle_increment = pinion_tooth_period / 2;
```

```
      wheel_anim_rot = wheel_frame_rotation * (frame_loop - 1);
      pinion_anim_rot = pinion_frame_rotation * (frame_loop - 1);

      fprintf(frame_log,"frame %d: ",frame_loop);
      fprintf(frame_log,"wheel rotation: %5.5f ",wheel_anim_rot);
      fprintf(frame_log,"pinion rotation: %5.5f\n",pinion_anim_rot);

      overall_wheel_rotation = 45 + wheel_anim_rot;
      overall_pinion_rotation = 45 + pinion_angle_increment - pinion_anim_rot;
                       # so that wheel and pinion mesh
                       # correctly, regardless of number of
                       # teeth.

      pcr = wheel_pitch_circle_radius;
      draw_wheel_teeth(frame_file,wheel_teeth,overall_wheel_rotation,pcr);
      pcr = pinion_pitch_circle_radius;
      draw_pinion_teeth(frame_file,pinion_teeth,overall_pinion_rotation,pcr);
      fprintf(frame_file,"</svg>\n");
      fclose(frame_file);
   endfor
   fclose(converter_script);
   fclose(frame_log);
endif

if (no_mesh || single_mesh)
   fprintf(wp_file,"</svg>\n");
   fclose(wp_file);
endif
```

## Test run output (to *stdout*) from 16/6 test run:

**(This is the output, to *stdout* under Linux, of the Octave code when run with 16 wheel teeth and 6 pinion teeth).**

```
GNU Octave, version 3.6.2
Copyright (C) 2012 John W. Eaton and others.
This is free software; see the source code for copying conditions.
There is ABSOLUTELY NO WARRANTY; not even for MERCHANTABILITY or
FITNESS FOR A PARTICULAR PURPOSE.  For details, type `warranty'.

Octave was configured for "i486-pc-linux-gnu".

Additional information about Octave is available at
http://www.octave.org.

Please contribute if you find this software useful.
For more information, visit http://www.octave.org/help-wanted.html

Read http://www.octave.org/bugs.html to learn how to submit bug
reports.

For information about changes from previous versions, type `news'.

Number of wheel teeth ->16
Number of pinion teeth ->6
```

```
Simulate meshing? (0 for no, 1 for single mesh, 2 for animation)?
Simulate meshing? ->1
Clip dedenda to inner circle? (0 for no, 1 for yes)?
Clip dedenda? ->1
SVG output height (default 500) ->750
SVG output width (default 500) ->750
Will do single mesh simulation.
Wheel teeth: 16, Pinion teeth: 6
Wheel PCR: 8.00000
Pinion PCR: 3.00000

Wheel GCR: 1.50000
Pinion GCR: 4.00000

Wheel GC one rev degrees: 67.50000
Pinion GC one rev degrees: 480.00000
pinion GC one rev degrees > 360, so cutting back to 360
(Clearly optimum pinion GC angle should be within one rev
Wheel RR: 0.18750
Pinion RR: 1.33333

Wheel tooth tip angle:
    Angle in degrees: 5.62500
    Angle in radians: 0.09817
Pinion tooth tip angle
    Angle in degrees: 15.00000
    Angle in radians: 0.26180
Tooth tip gradiants:
Wheel TTG: 0.09849
Pinion TTG: 0.26795

Finding wheel GC angle
angle in degrees: 0.00000, grad_diff: -0.09849
angle in degrees: 1.00000, grad_diff: -0.09855
angle in degrees: 2.00000, grad_diff: -0.09854
angle in degrees: 3.00000, grad_diff: -0.09823
angle in degrees: 4.00000, grad_diff: -0.09737
angle in degrees: 5.00000, grad_diff: -0.09573
angle in degrees: 6.00000, grad_diff: -0.09307
angle in degrees: 7.00000, grad_diff: -0.08919
angle in degrees: 8.00000, grad_diff: -0.08387
angle in degrees: 9.00000, grad_diff: -0.07695
angle in degrees: 10.00000, grad_diff: -0.06824
angle in degrees: 11.00000, grad_diff: -0.05762
angle in degrees: 12.00000, grad_diff: -0.04497
angle in degrees: 13.00000, grad_diff: -0.03019
angle in degrees: 14.00000, grad_diff: -0.01322
angle in degrees: 15.00000, grad_diff: 0.00598
Finding pinion GC angle:
angle in degrees: 0.00000, grad_diff: -0.26795
angle in degrees: 1.00000, grad_diff: -0.26802
angle in degrees: 2.00000, grad_diff: -0.26820
```

```
angle in degrees: 3.00000, grad_diff: -0.26848
angle in degrees: 4.00000, grad_diff: -0.26882
angle in degrees: 5.00000, grad_diff: -0.26920
angle in degrees: 6.00000, grad_diff: -0.26959
angle in degrees: 7.00000, grad_diff: -0.26997
angle in degrees: 8.00000, grad_diff: -0.27030
angle in degrees: 9.00000, grad_diff: -0.27057
angle in degrees: 10.00000, grad_diff: -0.27075
angle in degrees: 11.00000, grad_diff: -0.27080
angle in degrees: 12.00000, grad_diff: -0.27071
angle in degrees: 13.00000, grad_diff: -0.27045
angle in degrees: 14.00000, grad_diff: -0.26999
angle in degrees: 15.00000, grad_diff: -0.26930
angle in degrees: 16.00000, grad_diff: -0.26836
angle in degrees: 17.00000, grad_diff: -0.26714
angle in degrees: 18.00000, grad_diff: -0.26562
angle in degrees: 19.00000, grad_diff: -0.26377
angle in degrees: 20.00000, grad_diff: -0.26158
angle in degrees: 21.00000, grad_diff: -0.25900
angle in degrees: 22.00000, grad_diff: -0.25603
angle in degrees: 23.00000, grad_diff: -0.25263
angle in degrees: 24.00000, grad_diff: -0.24878
angle in degrees: 25.00000, grad_diff: -0.24447
angle in degrees: 26.00000, grad_diff: -0.23966
angle in degrees: 27.00000, grad_diff: -0.23434
angle in degrees: 28.00000, grad_diff: -0.22849
angle in degrees: 29.00000, grad_diff: -0.22208
angle in degrees: 30.00000, grad_diff: -0.21510
angle in degrees: 31.00000, grad_diff: -0.20752
angle in degrees: 32.00000, grad_diff: -0.19934
angle in degrees: 33.00000, grad_diff: -0.19052
angle in degrees: 34.00000, grad_diff: -0.18106
angle in degrees: 35.00000, grad_diff: -0.17093
angle in degrees: 36.00000, grad_diff: -0.16012
angle in degrees: 37.00000, grad_diff: -0.14863
angle in degrees: 38.00000, grad_diff: -0.13642
angle in degrees: 39.00000, grad_diff: -0.12349
angle in degrees: 40.00000, grad_diff: -0.10984
angle in degrees: 41.00000, grad_diff: -0.09543
angle in degrees: 42.00000, grad_diff: -0.08028
angle in degrees: 43.00000, grad_diff: -0.06436
angle in degrees: 44.00000, grad_diff: -0.04767
angle in degrees: 45.00000, grad_diff: -0.03019
angle in degrees: 46.00000, grad_diff: -0.01193
angle in degrees: 47.00000, grad_diff: 0.00712
First approximation of wheel GC angle for addendum tip
     Angle in degrees: 15.00000
     Angle in radians: 0.26180
First approximation of pinion GC angle for addendum tip
     Angle in degrees: 47.00000
     Angle in radians: 0.82030
```

```
Now starting Newton-Raphson iteration for wheel GC

After iteration step 16, angle is:
      Angle in degrees: 14.70086
      Angle in radians: 0.25658
After iteration step 17, angle is:
      Angle in degrees: 14.70086
      Angle in radians: 0.25658
After iteration step 18, angle is:
      Angle in degrees: 14.70086
      Angle in radians: 0.25658
After iteration step 19, angle is:
      Angle in degrees: 14.70086
      Angle in radians: 0.25658
After iteration step 20, angle is:
      Angle in degrees: 14.70086
      Angle in radians: 0.25658

Now starting NR iteration for pinion GC

After iteration step 16, angle is:
      Angle in degrees: 46.63118
      Angle in radians: 0.81387
After iteration step 17, angle is:
      Angle in degrees: 46.63118
      Angle in radians: 0.81387
After iteration step 18, angle is:
      Angle in degrees: 46.63118
      Angle in radians: 0.81387
After iteration step 19, angle is:
      Angle in degrees: 46.63118
      Angle in radians: 0.81387
After iteration step 20, angle is:
      Angle in degrees: 46.63118
      Angle in radians: 0.81387
Wheel addendum tip height: 9.31512
Pinion addendum tip height:     4.37179
Wheel bound radius: 10.31512
Pinion bound radius: 5.37179
leftmost x: -10.31512
bottommost y: -10.31512
rightmost x: 13.14997
topmost y: 13.14997
unscaled height: 23.46509
unscaled width: 23.46509
svg height: 750
svg width: 750
vertical scale factor: 31.96237
horiz scale factor: 31.96237
smaller scale factor: 31.96237
```

## Test run output picture from 16/6 test run:

(This is the SVG output of the Octave code when run under Linux with 16 wheel teeth and 6 pinion teeth).

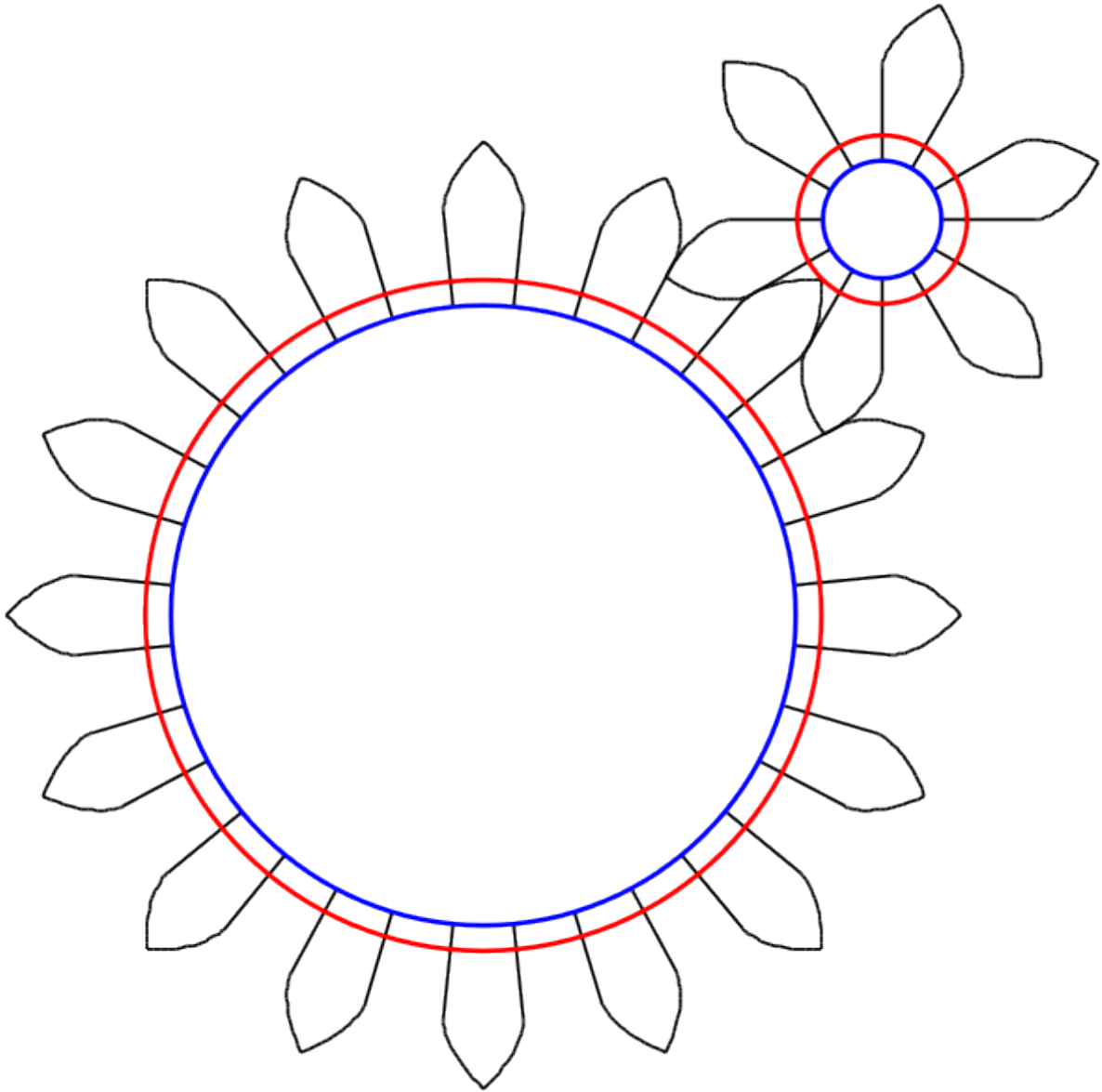

## Caveats, and Other Issues, with Use of this Octave Code:

1) I am happy to supply a raw copy of the Octave code (to save the reader from having to extract it from this document) on request.
I will also put a raw copy of the Octave code on the Internet (probably on github.org, if I understand the purpose of github.org correctly) if there is sufficient interest in this.

2) Selecting option 2 (animation) in response to the "simulate meshing" prompt produces a sequence of output files:

    file00001.svg
    file00002.svg
    file00003.svg

and so on (one for each frame of the animation).  It also produces a shell script (convert_script: you need to make it executable with chmod +x before it will run; sorry, I didn't have the stamina to implement this within the Octave code) for converting the frames from svg to png (this requires Imagemagick, reference [11], to be installed).

The whole animation can then be viewed with the Imagemagick *animate* command:

    animate *.png

or converted to an animated GIF with a command like (also using Imagemagick):

    convert *.png mesh.gif

(I don't have much experience of this latter option of producing animated GIFs in this way; I have much more experience with the former option of viewing the PNG images directly with the *animate* command, so I am not particularly confident of the quality of the animated GIFs produced in the above way).

3) The outer (red) circle on the wheel is the wheel dedendum base circle that the pinion tooth tips *just touch*.  Likewise, the outer (red) circle on the pinion is the pinion dedendum base circle that the wheel tooth tips *just touch*.

Clearly physically using these red circles for the dedendum base circles would almost certainly cause unwanted contact with the addendum tips (I think that *interference* is the term used in mechanical engineering), and so would be highly undesirable.

Hence, we need to allow some clearance to prevent this from happening.  The inner (blue) circles on the wheel and pinion are versions of the outer (red) circles with the radii reduced (by half a module, it appears from the code; it is a while since I wrote the code) in order to allow for some clearance.  It would probably be better to allow the user to set this clearance; I intend to make this clearance user-settable when I release the next version of this code.

4) Answering "1" to "Clip dedenda to inner circle?" gets rid of the "nexus of darkness" (extension of the wheel and pinion dedenda to the centres of the wheel and pinion) in the middle of the gears, which a programmer here in Auckland who saw one of my gear meshing animations in 2014 regarded as being a revolting eyesore. (I am happy to supply a picture of a gear drawn with the "nexus of darkness" included if the reader wants). He coined the phrase "nexus of darkness" to describe it. I intend to ask him if he is happy for me to acknowledge him (with his full name, which I currently do not know) for this in my next article on arXiv.

5) The programmer in question liked the result of my eliminating the "nexus of darkness". He said that the resulting animation was "soothing" and should be used as a screen saver!

6) **WARNING: this is a known problem with the code:** Certain combinations of number of wheel teeth and number of pinion teeth can cause the wheel and pinion teeth to overlap (in the SVG output picture) rather than meshing correctly when option 1 (for single mesh), and perhaps also option 2 (for animation) is selected.

However, I believe that the shapes of the wheel and pinion themselves are still correct in this case. One combination of number of wheel and pinion teeth that I know to cause this problem is 16 wheel teeth and 7 pinion teeth. I presume that the problem is related to the combination of parity (oddness or evenness) of the number of wheel teeth and parity of the number of pinion teeth. Clearly this is a problem that needs to be fixed, but I currently have higher-priority problems to fix with this code than this one.

7) **WARNING: this is a known potential problem with the code:** Since I don't have any proof of the convergence, in general, of the Newton-Raphson iteration (just empirical evidence that it converges, as you can see in the console output: thanks to Dr Radu Nicolescu for implicitly suggesting that I explicitly mention this point), I am not at all confident of the ability of this code to run correctly if the number of wheel teeth, or number of pinion teeth, is large.

## Acknowledgments:

1) Thanks to Ed Church and Andrew Dixon, programmers here in Auckland, for inspiring my interest in 3D printing, and giving me access to a 3D printer, back in late 2013.

2) Thanks also to Andrew Dixon for 3D printing (and paying for the filament himself, if I rememeber rightly) a pair of gears output by this Octave code in 2014.

3) Thanks to Dr Radu Nicolescu of the University of Auckland Computer Science Department for taking an interest in this project, and offering to look at this paper once it appears on arXiv.

4) Thanks to Mrs Adriana Ferraro, also of the University of Auckland Computer Science Department for taking an interest in this project.

5) Thanks to Dr Dmitry Berdinsky, formerly of the University of Auckland Computer Science Department (now a university lecturer in Thailand, I believe) for taking an interest in this project, and offering to look at this paper once it appears on arXiv.

6) Thanks to the programmer in Auckland (whose full name I do not yet know, but whom I intend to try to contact) who coined the term "nexus of darkness" (as discussed above), advised me to eliminate it, and described the result of eliminating it as being "soothing".

7) Thanks to Mr Chris Chitty of Massey University (School of Engineering and Advanced Technology, I believe, in Albany) for taking an interest in my solar-thermal power project in general, and suggesting a possible solution to another problem associated with it.

8) Thanks to Mr Ralph Versteegen, a PhD student in the University of Auckland Computer Science Department, for taking an interest in my solar-thermal power project in general, and discussing solutions to other problems associated with it.

9) Thanks to Dr Charles Elachi, director of NASA's Jet Propulsion Laboratory (reference [12]) for giving the public lecture in 2014 at which (as I discussed earlier) the solution to the tooth tip problem suddenly became clear to me.

10) Thanks to Associate Professor Nevil Brownlee of the University of Auckland Computer Science Department for taking a slight interest in this project, and giving me some advice on other aspects of solar-thermal power, particularly the possible use of Abner Doble's (1930s era, I believe) steam engines, which were apparently used by Doble in his steam-powered cars at the time. (Reference [13], which I *have not* read, or reference [14], which I have viewed briefly). If I remember rightly, Abner Doble claimed his steam engines to have a thermal power input of only about 100 or 200 kilowatts (much smaller than the smallest steam turbine engines in common use today, as far as I know), and a thermodynamic efficency of about 18 percent.

## **References:**

[1]   http://www.ivanpahsolar.com

[2]   http://www.openscad.org/

[3]   http://www.geartechnology.com/blog/dr-faydor-l-litvin-100-years-a-genius/

[4]   *Gear Geometry and Applied Theory*, second ed., F.L. Litvin, A. Fuentes. Cambridge: Cambridge University Press, 2004.

[5]   http://ntrs.nasa.gov/archive/nasa/casi.ntrs.nasa.gov/19900010277.pdf

[6]   http://www.csparks.com/watchmaking/CycloidalGears/index.jxl

[7]   Weisstein, Eric W. "Epicycloid." From *MathWorld*--A Wolfram Web Resource. http://mathworld.wolfram.com/Epicycloid.html

[8]   Weisstein, Eric W. "Hypocycloid." From *MathWorld*--A Wolfram Web Resource. http://mathworld.wolfram.com/Hypocycloid.html

[9]   https://www.ipenz.nz/home/news-and-publications/news-article/2015/09/10/2014-pickering-lecture-mars

[10]  https://www.haskell.org/hugs/

[11]  http://www.imagemagick.org/script/index.php

[12]  http://nmp.jpl.nasa.gov/bios/elachi/

[13]  Fox Stephen (1998). *The Strange Triumph of Abner Doble,* Invention & Technology Magazine, Volume 14; Issue 1

[14]  http://www.svvs.org/doble.shtml